\documentclass[pra,aps,twocolumn,
notitlepage,longbibliography]{revtex4-1}
\usepackage[colorlinks=true, citecolor=red, urlcolor=blue ]{hyperref}
\usepackage{graphicx}
\usepackage{dcolumn}
\usepackage{bm}
\usepackage{amsmath,amsfonts}
\usepackage[english]{babel}
\usepackage{amsthm}
\usepackage{hyperref}
\usepackage{xcolor}
\usepackage{braket}
\theoremstyle{plain}
\usepackage{color}
\usepackage{amssymb}
\usepackage{amsthm}
\usepackage{amsfonts}
\usepackage{float}
\usepackage{graphicx}
\usepackage[format = plain,labelfont = bf,up, textfont = normal , up, justification =raggedright, singlelinecheck =false]{caption}
\usepackage{subcaption}
\usepackage{tabularx}
\DeclareMathOperator{\tr}{Tr}
\usepackage{graphicx}

\usepackage{color,soul}
\usepackage{amsmath}
\usepackage{braket}
\usepackage{latexsym}
\usepackage{bm}
\usepackage{graphics,epstopdf}

\usepackage{enumitem}
\usepackage{fmtcount}
\usepackage{booktabs}
\usepackage{csquotes}
\usepackage{epsfig}
\newcommand{\delete}[1]{}

\newtheorem{theorem}{Theorem}
\newtheorem{corollary}{Corollary}[theorem]

\usepackage{epsfig}
\def\tr{\text{tr}}
\theoremstyle{plain}

\begin{document}
\title{Closing loopholes of\\ measurement-device-independent nonlinear entanglement witnesses}

\author{Kornikar Sen, Chirag Srivastava, Ujjwal Sen}
\affiliation{Harish-Chandra Research Institute, A CI of Homi Bhabha National Institute, Chhatnag Road, Jhunsi, Allahabad 211 019, India}

\begin{abstract}
The concept of entanglement witnesses form a useful technique to detect entanglement in realistic quantum devices.  
Measurement-device-independent nonlinear entanglement witnesses (MDI-NEWs) are a kind of entanglement witnesses which eliminate  dependence on the correct alignments of measurement devices for guaranteeing the existence of entanglement and also detect more entangled states than their linear counterparts. While this method guarantees entanglement independent of measurement alignments, they are still prone to serving wrong results due to other loopholes. 
Here we study the response of MDI-NEWs to two categories of faults occurring in  experiments. In the first category, the detection loophole, characterized by lost and additional events of outcomes of measurements, is investigated, and bounds which guarantee entanglement  are obtained in terms of the efficiency of measurement being performed. In the second category, we study noise associated with the sets of additional quantum inputs required in  MDI-NEW scenarios. In this case, a type of noise is identified which still allows the MDI-NEWs to guarantee entanglement. We also show that  MDI-NEWs are less or equally robust in comparison to their linear counterparts under the same noise in additional quantum inputs, although the former group detects a larger volume of entangled states in the noiseless scenario than their linear cousins.
\end{abstract}
\maketitle

\section{Introduction}
Entanglement~\cite{horodecki09,Guhne09, das2017separability} 
is one of the most interesting and useful characteristics of quantum states. 
After its first recognition as a 
property of shared 
physical entities~\cite{EPR}, entanglement has received a significant amount of attention. 
In particular, entanglement acts as a resource in a variety of quantum mechanical protocols, including  quantum teleportation~\cite{PhysRevLett.70.1895}, quantum dense coding~\cite{PhysRevLett.69.2881}, and entanglement-based quantum cryptography~\cite{PhysRevLett.67.661}.


Consequently, various methods for detection of entanglement have been discovered and analyzed. The range criterion~\cite{HORODECKI1997333, PhysRevLett.80.5239}, positive partial transpose (PPT) criterion~\cite{PPT1,PPT2}, entropic criterion~\cite{HORODECKI1994147, PhysRevLett.86.5184}, entanglement witnesses (EWs)~\cite{PPT2, TERHAL2000319, doi:10.1080/09500340110105975}
 are some  examples. Among these different methods of detecting entanglement, (linear) entanglement witnesses have received a lot of attention, due to it being easier to implement in experiments than the other methods, and its usability in cases in which the state is only partially unknown. 
 Since the set of separable (unentangled) quantum states of shared systems is convex and closed, one can invoke the Hahn-Banach theorem~\cite{hahn1,hahn2} and be always able to draw a hyperplane to ``separate'' any entangled state 
 from the set of all separable states. 
 This is the idea in the concept of linear entanglement witnesses, and the expectation value of a linear operator 
 acts as the hyperplane for the detection of entangled states. 
Entanglement witnesses have been utilized for detecting entanglement in many experiments~\cite{exp1,exp2,exp3,exp4,exp5,exp6}.

The instruments available in laboratories do not operate ideally and are not perfectly efficient. Thus,
the process of implementing an EW in real situations face various imperfections, for example, the ``wrong measurement loophole'' and the ``detection loophole''~\cite{WM5,WM6,WM4,WM3,WM2,WM1}. If the measurement settings get altered during the evaluation of the expectation value of an EW, then due to the incorrectness in the obtained information, a separable state may appear as entangled. This is called the wrong measurement loophole. Similar complications may also happen due to detection loophole, where one or more events get lost or falsely appear in the measurement procedure. In the context of Bell inequality, the significance of the detection loophole has been discussed in some detail~\cite{Bell_theory1,Bell_theory2,Bell_theory3,Bell_theory4,Bell_theory5,Bell_theory6,Bell_theory7,Bell_theory8,Bell_theory9}, and the related experiments performed include~\cite{Bell_exp1,Bell_exp2,Bell_exp3,Bell_exp4,Bell_exp5,Bell_exp6}. 
To avoid the wrong measurement loophole present in the detection of entangled states via entanglement witnesses, measurement-device-independent entanglement witnesses (MDI-EWs) were introduced in Ref.~\cite{MDI1} based on a semi-quantum non-local game~\cite{Buscemi12}. Nevertheless, the detection loophole was present in MDI-EW. The effect of the detection loophole, and the path for bypassing it, was analyzed in~\cite{MDI2}. Further works on 
MDI-EWs include~\cite{MDI3,MDI4,MDI5,MDI6}.

Detection of entangled states using MDI-EWs requires additional quantum states as inputs. From therein arises the ``noisy quantum input loophole'' in the process of detection of entangled states using MDI-EWs~\cite{MDI6}, that is, when the input states themselves become faulty. 

There is as yet no efficient way to detect all the entangled states of a shared Hilbert space of an arbitrary dimension. This is the well-known separability problem. The method of detecting entanglement by linear entanglement witnesses is also not an exception: no single linear EW can detect all entangled states. However, one can construct better entanglement witnesses than the linear ones by ``bending'' the linear EWs towards negativity so that it still has semi-positive values for all separable states. Such entanglement witnesses, due to their nonlinear nature, are called nonlinear entanglement witnesses (NEWs)~\cite{NL1,amar,NL2,NL3,NL4,NL5,NL6,NL7,NL8,NL9,NL10,NL11,NL12,NL13,NL14,NL15}. Nonlinear entanglement witnesses, due to their  construction, can detect more entangled states than their linear counterparts. However, the construction of a nonlinear witness, which can detect all the entangled states, that is, the (necessarily nonlinear) operator forming the boundary surface that separates the set of separable states from the set of entangled states, which will then solve the separability problem, is still open. The method of detecting entangled states using nonlinear witnesses also encounters the wrong measurement and detection loopholes. It was recently shown that  measurement-device-independent nonlinear entanglement witnesses~(MDI-NEWs) that are better than MDI-EWs can also be constructed~\cite{amar2}.  

In this article, we 
explore the effects of detection  and noisy quantum input loopholes in MDI-NEWs.
In case of the detection loophole, we consider three different scenarios, with the first and second being those of 
additional events but no lost events in the detection process, and the opposite.
The third is the most general scenario, where events can get lost as well as there can be unwanted excess events. Considering these three scenarios separately, we determine modified conditions on the MDI-NEWs such that no false occurrence of entangled states happens in any of these circumstances. We next show the MDI-NEWs never depict a separable state as entangled even in the presence of certain types of noise in the set of input states. Strangely, we realize that if due to any noise in input states, MDI-EWs provide an incorrect result, then the corresponding MDI-NEWs will do the same. 

The rest of the paper is organized as follows: in Sec.~\ref{sec2}, we will recapitulate formal definitions of EWs, MDI-EWs, and MDI-NEWs, which will also help to 
set the notations. In Sec.~\ref{sec3}, we will discuss how the detection loophole can have an impact on MDI-NEWs and how it can be avoided if the apparatus's efficiencies are known. The consequence of noisy quantum input loophole will be discussed in Sec.~\ref{sec5}. We present a conclusion in Sec.~\ref{sec6}.

\label{sec1}
\section{Prerequisites}
\label{sec2}

In this section, we will briefly review some topics which will be needed in the remaining part of the paper.

\subsection{Entanglement witnesses}
\label{sec2A}
Consider a composite Hilbert space, $\mathcal{H}_A\otimes\mathcal{H}_B$. The concept of linear entanglement witnesses consists in choosing 
a hermitian operator, $W$, acting on $\mathcal{H}_A\otimes\mathcal{H}_B$ such that $\tr(W\sigma)\geq0$ for all separable states, \(\sigma\), on \(\mathcal{H}_A\otimes\mathcal{H}_B\), and $\tr(W\rho)<0$ for at least one entangled state, \(\rho\) on \(\mathcal{H}_A\otimes\mathcal{H}_B\). Therefore if we know that for a hermitian operator, \(W\),
\(\tr(W\sigma)\geq0\) for all separable states, \(\sigma\), and find that \(\tr(W\rho)<0\) for a given \(\rho\), then the state \(\rho\) must be entangled.
%
%
For example, consider a non-positive partial transpose (NPT) bipartite state,  $\tilde{\rho}$, acting on the Hilbert space $\mathcal{H}_A\otimes\mathcal{H}_B$. According to the definition of NPT states, the operator $\tilde{\rho}^{T_B}$ will have at least one negative eigenvalue, where $T_B$ denotes transposition on the second Hilbert space, $\mathcal{H}_B$. An NPT state is always an entangled state~\cite{PPT1,PPT2}.
Let us denote the eigenvector corresponding to any one of the negative eigenvalues of $\tilde{\rho}^{T_B}$ by $\ket{\tilde{\phi}}$. Then the operator $W_{\tilde{\phi}}=(\ket{\tilde{\phi}}\bra{\tilde{\phi}})^{T_B}$ is a valid linear entanglement witness operator which can successfully detect the entangled state $\tilde{\rho}$~\cite{Guhne09}.
\vspace{2mm} 
\\ 
\emph{Measurement-device-independent entanglement witnesses:}
Let us now briefly discuss the concept of 
MDI-EWs. Any entanglement witness operator, $W$, can be decomposed in terms of density matrices 
as
\begin{equation}
    W=\sum_{s,t}\alpha_{st}\tau_s^T\otimes\omega_t^T,\label{eq2}
\end{equation}
where $\tau_s$ and $\omega_t$ are  sets of density matrices, acting on the individual Hilbert spaces, $\mathcal{H}_A$ and $\mathcal{H}_B$, respectively, and $\alpha_{st}$ are real numbers. Consider a particular scenario where Alice and Bob share the state $\rho$, which acts on the same composite Hilbert space, $\mathcal{H}_A\otimes\mathcal{H}_B$. And they want to find out if the state $\rho$ is entangled. The states, $\tau_s$ ($\omega_t$), are available to Alice (Bob), which are called the \emph{input states}. Now, Alice (Bob) applies a  positive operator-valued measurement (POVM) on the state $\tau_s$ ($\omega_t$) and her (his) part of the state $\rho$. Each of the POVMs on Alice's and Bob's sides has only two distinct outcomes, say 0 and 1. Let the POVM operators corresponding to the outcome 1 be $A_1$ and $B_1$ for Alice's and Bob's measurements, respectively. The probability that each of them will get outcome 1 when the input states utilized are $\tau_s$ and $\omega_t$, 
is given by
\begin{equation*}
    P_{11}^{st}(\rho)=\tr[(\tau_s\otimes\rho\otimes\omega_t)(A_1\otimes B_1)].\label{eq12}
\end{equation*}
The MDI-EW, $I_\alpha$, is thus defined as~\cite{MDI1}
\begin{equation}
    I_\alpha(P_\rho)=\sum_{s,t}\alpha_{st}P^{st}_{11}(\rho), \label{eq1}
\end{equation}
so that 
$I_\alpha(P_\sigma)\geq 0$ for all separable states, $\sigma$, and for any  dichotomic POVMs. Since the non-negativity of \(I_\alpha(P_\sigma)\) is independent of the choice of POVMs, the witness is called measurement-device independent. On the other hand, when $A_1$ and $B_1$ are taken to be the maximally entangled states, i.e., $\ket{\Phi_A}\bra{\Phi_A}=\frac{1}{d_A}\sum_{i,j} \ket{ii}\bra{jj}$ and $\ket{\Phi_B}\bra{\Phi_B}=\frac{1}{d_B}\sum_{i,j} \ket{ii}\bra{jj}$ respectively, Eq. \eqref{eq1} reduces to $I_\alpha(P_\rho)=\frac{\tr(W\rho)}{d_Ad_B}$, 
where $d_A$ and $d_B$ are respectively the dimensions of $\mathcal{H}_A$ and $\mathcal{H}_B$. Thus, for this particular set of POVMs, the MDI-EW can detect all the states that the usual measurement-device-dependent EW, $W$, can detect.


\subsection{Nonlinear entanglement witnesses}
Consider the witness operator $W_{\tilde{\phi}}$, an arbitrary state $\ket{\psi}$ acting on $\mathcal{H}_A\otimes\mathcal{H}_B$, and a corresponding operator $X=\ket{\tilde{\phi}}\bra{\psi}$. Then using these operators and states, a nonlinear operator can be constructed as~\cite{NL1}
\begin{eqnarray*}
    F_{\tilde{\phi}}(\rho)&=&\braket{W_{\tilde{\phi}}}-\frac{1}{s(X)}\left(\left< X^{T_B}\right>\left<{\left(X^{T_B}\right)^\dagger}\right> \right),
\end{eqnarray*}
where the expectation values are taken over the state $\rho$. 
We denote the square of the largest Schmidt decomposition coefficient of $\ket{\psi}$ by $s(X)$. The operator, $F_{\tilde{\phi}}(\sigma)$, is non-negative for all separable states, $\sigma$. But the second term in $F_{\tilde{\phi}}$ will always have a non-positive value, so that $F_{\tilde{\phi}}(\rho)\leq W_{\tilde{\phi}}$. Therefore, $F_{\tilde{\phi}}(\sigma)$ can be defined as a nonlinear entanglement witness that can detect more entangled states than \(W_{\tilde{\phi}}\)~\cite{NL1}. \vspace{2mm}\\
\emph{Measurement-device-independent nonlinear entanglement witnesses:}
It is also possible to design nonlinear witnesses in a measurement device-independent way~\cite{amar2}. The basic detection procedure is the same as the MDI-EWs, with the only difference being that in this case, Alice and Bob require the maximally mixed states, $m_A$ and $m_B$, also as inputs. We denote the probability of obtaining the outcomes $a$ and $b$ when the ``input states'' are $m_A$ and $m_B$ as $P^{\mathcal{A}\mathcal{B}}_{ab}$.

Any operator can be decomposed as a sum of hermitian and anti-hermitian operators. Thus we can write $X^{T_B}=H_1+iH_2$, where $H_1$ and $H_2$ are hermitian operators. Again both of the hermitian operators, $H_1$ and $H_2$, can be decomposed in terms of the same local density matrices, $\tau_s$ and $\omega_t$, as in Eq.~(\ref{eq2}), as 
\begin{equation}
      H_1=\sum_{s,t}\beta_{st}\tau_s^T\otimes\omega_t^T,\text{ } H_2=\sum_{s,t}\gamma_{st}\tau_s^T\otimes\omega_t^T.\label{eq3}
\end{equation}
An MDI-NEW $N_{\tilde{\phi}}$ can be defined, in terms of the decomposition coefficients of the hermitian operators, $W_{\tilde{\phi}}$, $H_1$, and $H_2$, given in Eqs.~\eqref{eq2} and~\eqref{eq3}, as
 
\begin{widetext}
\begin{equation}
    N_{\tilde{\phi}}(P)=I_\alpha(P)-\frac{1}{s(X)d_Ad_BP_{11}^{\mathcal{AB}}}
    \left[ \left(\sum_{s,t} \beta_{st} P_{11}^{st}\right)^2+\left(\sum_{s,t} \gamma_{st} P_{11}^{st}\right)^2\right].\label{eq4}
\end{equation}
Let Alice and Bob share a separable state, $\sigma=\sum_i p_i \sigma_A^i\otimes\sigma^i_B$.
In such a situation Eq.~\eqref{eq4} reduces to
\begin{equation}
 N_{\tilde{\phi}}(P_\sigma)=\tr{\left[\sum_i p_iA_1^i\otimes B_1^i W_{\tilde{\phi}}\right]}-\frac{1}{s(X)\tr\left[\sum_i p_i A_1^i\otimes B_1^i\right]}
 \left[\left( \tr\left[\sum_i p_iA_1^i\otimes B_1^i H_1\right]\right)^2
 +\left(\tr\left[\sum_i p_iA_1^i\otimes B_1^i H_2\right]\right)^2\right].
\end{equation}
\end{widetext}
Here $A_1^i=\left(\tr_A{[A_1(\mathbb{I}_A\otimes\sigma^i_A)]}\right)^T$ and $B_1^i=\left(\tr_B{[B_1(\sigma_B^i\otimes\mathbb{I}_B)]}\right)^T$. Thus it can be seen that $N_{\tilde{\phi}}(P_\sigma)=T_QF_{\tilde{\phi}}(Q)$, 
where $T_Q=\sum_i p_i \tr\left[{ A_1^i\otimes B_1^i}\right]$ and $Q= \sum_i p_i A_1^i\otimes B_1^i/T_Q$. Since  
$A_1^i$ and $B_1^i$ are effective POVMs acting on Alice's and Bob's Hilbert spaces respectively and $Q$ is a separable state, we have $T_Q\geq 0$ and $F_{\tilde{\phi}}(Q)\geq 0$.
Thus it confirms that the nonlinear witness function, $N_{\tilde{\phi}}$,  is positive for any joint dichotomic measurement applied by Alice and Bob on separable states. 
If Alice and Bob choose the joint POVM operators, $A_1$ and $B_1$, to be maximally entangled states, then the expression in Eq. \eqref{eq4} reduces to $N_{\tilde{\phi}}(\rho)=\frac{F_{\tilde{\phi}}}{d_Ad_B}$. Therefore, in this case, $N_{\tilde{\phi}}$ is equally effective as $F_{\tilde{\phi}}$ for the detection of entangled states.

\section{Effects of detection loophole on entanglement detection via MDI-NEW}
\label{sec3}
 Even though the wrong measurement loophole can be ignored while using MDI-EWs, the detection loophole can still be present, as shown  in~\cite{MDI2}.
The detection loophole arises in experiments due to lost  or/and  additional events during the implementation of measurements.
 It may hamper the determination of  probabilities of outcomes of measurements, which in turn may cause a wrong certification of an entangled state.
 In this section, we want to determine how this loophole may affect the entangled state detection process using MDI-NEWs.

In presence of the detection loophole, the measured value of the probability of an outcome $(a,b)$, when the input states are $\tau_s$ and $\omega_t$, is $ (P_{ab}^{st})_m=\frac{ (n_{ab}^{st})_m}{ (N^{st})_m}$, where $ (n_{ab}^{st})_m$ and $ (N^{st})_m$ denote the number of times the experimentalist got the outcome $(a,b)$ and the total number of outcomes respectively. We can write 
 \begin{equation*}
     (P_{ab}^{st})_m=\frac{ ({n}_{ab}^{st})_i+\epsilon^{st}_{+ab}-\epsilon^{st}_{-ab}}{ \sum_{a,b}[({n}_{ab}^{st})_i+\epsilon^{st}_{+ab}-\epsilon^{st}_{-ab}]},
  \end{equation*}
 where $({n}_{ab}^{st})_i$, and $\epsilon^{st}_{+ab}$ and and $\epsilon^{st}_{-ab}$ denote the number of times the outcome $(a,b)$ should have clicked in the ideal case, and the corresponding number of additional and lost events respectively. A similar relation is also valid when the input states are $m_A$ and $m_B$.
For demonstration, 
we assume that the number of additional or lost events do not depend on the input states, 
and also are independent of the output, so that we set
$\epsilon^{st}_{+ab}=\epsilon_+$ and $\epsilon^{st}_{-ab}=\epsilon_-$. Moreover, total number of outputs in the ideal case, $(N^{st})_{i}=\sum_{ab}\left(n^{st}_{ab}\right)_{i}$, when input states are $\tau_s$ and $\omega_t$, is also assumed to be fixed for all input states, and we set $(N^{st})_i=\bar{N}$.

Keeping in mind the two types of undesirable events, two corresponding efficiencies are defined: the ``additional event efficiency'', $\eta_+=\frac{\Bar{N}}{\Bar{N}+E_+}$, and the ``lost event efficiency'', $\eta_-=\frac{\Bar{N}-E_-}{\Bar{N}}$, where $E_\pm=\sum_{a,b} \epsilon_{\pm }$.

We will now consider the following three types of scenarios separately.
 
\noindent \textbf{Case 1:} In this case, we restrict the additional event efficiency to unity: $\eta_+=1$. But the lost event efficiency, $\eta_-\in [0,1]$.  
 
\noindent  \textbf{Case 2:} This is exactly opposite to the previous case, i.e., $\eta_+\in [0,1]$ whereas $\eta_-=1$.
 
\noindent \textbf{Case 3:} In this case, we consider the general situation. Here both the efficiencies are arbitrary, i.e., $\eta_+\in [0,1]$ and  $\eta_-\in [0,1]$.
 
 We want to find the modified bound 
 on the measured values of nonlinear witnesses, $(N_{\tilde{\phi}})_m$, so that no separable state show up as entangled.
In the following subsections, we will discuss the three cases individually and will find the corresponding bound on $(N_{\tilde{\phi}})_m$ for each case.  

\subsection{Arbitrary lost event efficiency}
Since Alice and Bob apply dichotomic POVMs, there are four possible outcomes, \textit{viz.} (0,0), (0,1), (1,0), (1,1), among which we are interested in the outcome (1,1) only. The measured value of the probability of an outcome (1,1), when the apparatus has unit additional event efficiency, is 
\begin{equation}
    (P_{11}^{st})_m=\frac{(n_{11}^{st})_i-\epsilon_{-}}{\sum_{a,b} (n_{ab}^{st})_i-4\epsilon_{-}}=\frac{(P_{11}^{st})_i}{\eta_-}-\frac{1-\eta_-}{4\eta_-}. \label{eq11}
\end{equation}
A similar relation is also true for $\left(P_{11}^{\mathcal{AB}}\right)_m$. We will be using 
the following notations:
\begin{equation*}
    (I_c)_{m/i}=\sum_{s,t} c_{s,t}(P_{11}^{st})_{m/i}\text{ and }K=s(X)d_Ad_B, 
\end{equation*}
where $c$ can be $\alpha$, $\beta$, or $\gamma$. The measured value of $I_c$ is given by
\begin{equation}
    (I_c)_m=\frac{(I_c)_i}{\eta_-}-\frac{1-\eta_-}{4\eta_-}\sum_{s,t}c_{st}=\frac{(I_c)_i}{\eta_-}+\mathcal{P}_c \label{eq8},
\end{equation}
 where $\mathcal{P}_c=\frac{\eta_--1}{4\eta_-}\sum_{s,t}c_{st}$. Therefore, the measured value of the nonlinear witness is 
\begin{equation}
    (N_{\tilde{\phi}})_m=(I_\alpha)_m-\frac{1}{K(P_{11}^{\mathcal{AB}})_m}[(I_\beta)_m^2+(I_\gamma)_m^2].\label{eq7}
\end{equation}

Using Eqs.~\eqref{eq8} and~\eqref{eq7}, we can get a relation between the measured and true values of the nonlinear witness operator: 
\begin{eqnarray*}
    (N_{\tilde{\phi}})_m=\frac{1}{\eta_-}\left[(N_{\tilde{\phi}})_i+\frac{(I_\beta)_i^2+(I_\gamma)_i^2}{K(P_{11}^{\mathcal{AB}})_i}\right]-\frac{(I_\beta)_m^2+(I_\gamma)_m^2}{K(P_{11}^{\mathcal{AB}})_m}\\+\mathcal{P}_\alpha.
\end{eqnarray*}
If the state is entangled and the corresponding ideal value of the nonlinear witness operator is negative, i.e. if $(N_{\tilde{\phi}})_i<0$, then the measured value of the operator will satisfy the 
inequality
\begin{eqnarray}
    (N_{\tilde{\phi}})_m<\frac{1}{\eta_-}\frac{(I_\beta)_i^2+(I_\gamma)_i^2}{K(P_{11}^{\mathcal{AB}})_i}
    -\frac{(I_\beta)_m^2+(I_\gamma)_m^2}{K(P_{11}^{\mathcal{AB}})_m}+\mathcal{P}_\alpha. \label{eq9}
\end{eqnarray}
Using Eqs.~\eqref{eq11},~\eqref{eq8}, and~\eqref{eq9}, we finally get the bound on $(N_{\tilde{{\phi}}})_m$ for detection of entangled states correctly:
\begin{widetext}
\begin{equation}
     (N_{\tilde{\phi}})_m<\frac{{\eta_-}}{K\left(\eta_-(P_{11}^{\mathcal{AB}})_m+\frac{1-\eta_-}{4}\right)}\left[\left\{(I_\beta)_m-\mathcal{P}_\beta\right\}^2+\left\{(I_\gamma)_m-\mathcal{P}_\gamma\right\}^2\right]-\frac{1}{K(P_{11}^{\mathcal{AB}})_m}\left\{(I_\beta)_m^2+(I_\gamma)_m^2\right\}+\mathcal{P}_\alpha. \label{eq10}
\end{equation}
\end{widetext}
This inequality reports that even if $(N_{\tilde{\phi}})_m<0$ for a bipartite state, unless the state satisfies~\eqref{eq10}, the state may not be entangled. For $\eta_-=1$ we have $\mathcal{P}_\alpha=\mathcal{P}_\beta=\mathcal{P}_\gamma=0$, and in that case, the inequality~\eqref{eq10} reduces to  
the condition for detection of entangled state in the ideal case.
\subsection{Arbitrary additional event efficiency}
Now we proceed to the next situation, where the additional event efficiency is non-unit but the lost event efficiency, $\eta_-=1$. In this case, the measured value of  probability of the output $(1,1)$, when the input states are fixed to be $\tau_s$ and $\omega_t$, is 
\begin{eqnarray*}
    (P_{11}^{st})_m=\frac{({n}_{11}^{st})_i+\epsilon_{+}}{\sum_{a,b}({n}_{ab}^{st})_i+4\epsilon_{+}}=\eta_+\left\{(P_{11}^{st})_i+\frac{1-\eta_+}{4\eta_+}\right\}.
\end{eqnarray*}
A similar transformation holds for the input states $m_A$ and $m_B$.
Therefore, in this case, we have
    $(I_c)_m=\eta_+(I_c)_i+\mathcal{Q}_c$, where $\mathcal{Q}_c=\frac{1-\eta_+}{4}\sum_{s,t}c_{s,t}.$
Following the same path of calculations as in the preceding case, we find that when the ideal value of the nonlinear witness operator, $(N_{\tilde{\phi}})_i$, is less than zero, the corresponding measured value of the operator will satisfy 
\begin{widetext}
\begin{equation*}
        (N_{\tilde{\phi}})_m < \frac{1}{K\left[(P_{11}^{\mathcal{AB}})_m-\frac{1-\eta_+}{4}\right]}\left[\{(I_\beta)_m-\mathcal{Q}_\beta\}^2+\{(I_\gamma)_m-\mathcal{Q}_\gamma\}^2\right]-\frac{1}{K(P_{11}^{\mathcal{AB}})_m}[(I_\beta)_m^2+(I_\gamma)_m^2]+\mathcal{Q}_\alpha.
\end{equation*}
\end{widetext}
\subsection{Arbitrary additional and lost event efficiency}
In this part, we consider the  general case, that is $\eta_{\pm}\leq 1$. Here, the measured value of $P_{11}^{st}$ is 
\begin{equation*}
    \left(P^{st}_{11}\right)_m=\left(\eta_-+\frac{1}{\eta_+}-1\right)^{-1}\left[\left(P^{st}_{11}\right)_t+\frac{1}{4}\left(\eta_-+\frac{1}{\eta_+}-2\right)\right].
\end{equation*}
In such a situation, the bound on the measured value of the nonlinear witness for detection of entangled states is given by
\begin{widetext}
\begin{equation*}
    (N_\phi)_m < \frac{1}{K\left[\left(P_{11}^{AB}\right)_m-\frac{1-C}{4}\right]}\left[\{(I_\beta)_m-\mathcal{R}_\beta\}^2+\{(I_\gamma)_m-\mathcal{R}_\gamma\}^2\right]-\frac{1}{K\left(P_{11}^{AB}\right)_m}[(I_\beta)_m^2+(I_\gamma)_m^2]+\mathcal{R}_\alpha,
\end{equation*}
\end{widetext}
where   $C=\left(\eta_-+\frac{1}{\eta_+}-1\right)^{-1}$ and
    $\mathcal{R}_c=\frac{\sum_{s,t}c_{s,t}}{4}(1-C).$
\section{Consequences of noisy quantum inputs of MDI-NEW}
\label{sec5}

The MDI variety of entanglement witnesses, while clearing away the wrong measurement loophole, actually brings in another loophole, the so-called ``noisy quantum input'' loophole~\cite{MDI6}. 
%
In realistic situations, the input states $\tau_s$, $\omega_t$, $m_A$, and $m_B$ can get affected by any noise present in the apparatuses or other parts of the environment. Such noise may not always be insignificant, and they can  affect the validity of the detection of entangled states. The noisy quantum inputs are  $\Lambda_{st}(\tau_s\otimes\omega_t)$ and $\Lambda_{AB}(m_A\otimes m_B)$, where $\Lambda$ is the noise map. An important class of noise is what is referred to as 
``uniform'' noise.
For any noise from that class, the noise map acts uniformly on all the inputs, i.e.,  $\Lambda_{st}$ ($\Lambda_{AB}$) does not have any dependence on $s$ or $t$ ($A$ or $B$).

There are situations when, due to uniform noise in the instruments and the environment, a lesser number of entangled states might get detected by the MDI-NEW, but no separable state will erroneously be pointed out
as entangled. In the following subsection, we will discuss  some implications of uniform noise on MDI-NEWs.
And in the subsection next to that, we provide a comparison between MDI-NEWs and MDI-EWs.

\subsection{When uniform noise is resilient to noisy quantum input loophole}
We want to identify a set of noise maps within the class of uniform noise maps that still guarantee  entanglement measurement-device independently, i.e., they still give semi-positive values for all separable states, measurement device independently. Let Alice and Bob share the separable state, $\sigma=\sum_i p_i \sigma^i_A\otimes \sigma^i_B$. The MDI-NEW for this state, with the  quantum inputs affected by the uniform noise $\Lambda$, is given by 
\begin{widetext}
\begin{equation*}
 N^\Lambda_{\tilde{\phi}}(P_\sigma)=\sum_{s,t}\alpha_{st}\tr{[G_1' \Lambda(\tau_s^{T}\otimes\omega_t^{T})]}-\frac{1}{s(X)\tr[G_1' \Lambda(\mathbb{I}_{AB}
 )]}\left[ \left(\sum_{s,t} \beta_{st}\tr[G_1' \Lambda(\tau_s^{T}\otimes\omega_t^{T})]\right)^2+\left(\sum_{s,t} \gamma_{st} \tr[G_1' \Lambda(\tau_s^{T}\otimes\omega_t^{T})]\right)^2\right],\end{equation*}
\end{widetext}
where $G_1'=\sum_i p_i A_1^i\otimes B_1^i$.

Let us denote the adjoint of the map $\Lambda$ by $\Lambda^+$. Then according to the definition of adjoint maps, $\tr{\left[\sum_i p_i A_1^i\otimes B_1^i \Lambda\left(\tau_s^{T}\otimes\omega_t^{T}\right)\right]}=\tr{\left[\Lambda^+\left(\sum_i p_i A_1^i\otimes B_1^i\right)\tau_s^{T}\otimes\omega_t^{T}\right]}$. Let us now consider the special case where the adjoint of the noise, maps the set of separable positive semidefinite operators to operators of the same set. We denote the noise maps from this set as $\boldsymbol{\Lambda}$ and thus,
\begin{equation}
    \boldsymbol{\Lambda}^+(\sum_i p_i A_1^i\otimes B_1^i)=\sum_k C_k'\otimes D_k',
\end{equation}
where $A_1'$, $B_1'$, $C_k'$, and $D_k'$ are positive semidefinite operators.
Then the expression of the nonlinear MDI-EW reduces to $$N^{\boldsymbol{\Lambda}}_{\tilde{\phi}}(P_\sigma)=T_{Q'}F_{\tilde{\phi}}(Q'),$$
where $Q'=\sum_k\frac{ C_k'\otimes D_k'}{T_{Q'}}$ and  $T_{Q'}=\sum_k\tr{[ C_k'\otimes D_k']}\geq 0$. Hence we have $N^{\boldsymbol{\Lambda}}_{\tilde{\phi}}(P_\sigma)\geq 0$ for all separable states. We can thus state the following result.
\begin{theorem}
 The MDI-NEW will never erroneously indicate any separable state as entangled if the adjoint of the uniform noisy map over the quantum inputs maps the set of separable positive semidefinite operators to  operators of the same set.
\end{theorem}

Consider a subset of uniform noise maps, \emph{viz.}, local uniform noise maps, such that for an arbitrary element of that subset, \(\Lambda_l\), we have 
\(\Lambda_l = \Lambda_1 \otimes \Lambda_2\), so that 
$\Lambda_l(\tau_s\otimes\omega_t)=\Lambda_1(\tau_s)\otimes\Lambda_2(\omega_t)$ and  $\Lambda_l(m_A\otimes m_B)=\Lambda_1(m_A)\otimes\Lambda_2(m_B)$. 
Notice that the map $\Lambda_l$ takes the set of separable  operators onto the same set. This leads us to the following corollary.
\begin{corollary}
The MDI-NEW will not indicate a separable state as entangled in the presence of local uniform noise if the adjoints of the local noise maps keep semipositivity. 
\end{corollary}
\subsection{Comparison of MDI-EW and MDI-NEW in presence of noise}
Experimental implementation of nonlinear MDI-EWs will require more measurements than for 
the corresponding linear ones. However, the former can detect more entangled states.

Another important question to ask is about the robustness of their performances under the influence of noise, and in this case we find that the MDI-NEW inherits a disconcert from the corresponding MDI-EW. Precisely, we have the following proposition in this regard.
\vspace{2mm}
\newline
\textbf{Proposition.}
\textit{For the noisy quantum inputs, if the MDI-EW mistakenly detects a separable state as entangled, then the corresponding MDI-NEW will provide the same erroneous result.}
\vspace{2mm}
\newline
\textbf{Proof:}
To construct the nonlinear witness given in Eq. \eqref{eq4}, we have subtracted a positive nonlinear term from the linear witness. It is therefore straightforward that the MDI-NEW's value will always be lower or equal to the value of the corresponding MDI-EW. Hence, if the MDI-EW has a negative value for any separable state for any noisy quantum inputs, then the corresponding MDI-NEW will also have a negative value for at least that separable state. 
\hfill \(\blacksquare\)
\vspace{2mm}
\newline
 \textbf{Remark:} This phenomenon can also be understood geometrically. We know that an MDI-EW can be represented by a hyperplane in the state space which never ``cuts'' the separable ball.
 To construct the MDI-NEW, given in Eq. \eqref{eq4},  the hyperplane of the corresponding linear witness is bent towards the separable ball. Thus, the MDI-NEW can be represented by a curved surface whose points lie either closer or equally distant from the separable ball when compared to the points on the hyperplane.
Similarly, for a given set of noisy quantum inputs, the noisy MDI-EW and the noisy MDI-NEW are represented by  another hyperplane and another (corresponding) curved surface in the state space, respectively. Again, for this noisy case, the curved surface is closer, than the hyperplane, to the separable ball.
 Hence, if due to imperfections, the separable ball cuts the noisy plane, representing the noisy MDI-EW, 
 the closer nonlinear surface, representing the noisy MDI-NEW, will cut at least equally. Thus, the MDI-EW is robust than the corresponding MDI-NEW under noisy inputs. To overturn this, the linear witness operator may be bent away from the separable ball, but then one has to give up on the advantage of MDI-NEWs over MDI-EWs, which is to detect a larger set of entangled states. 
\section{conclusion}
\label{sec6}
The technique of entanglement witnesses (EWs) constitutes an efficient way to detect  entangled quantum states in laboratories. Among the  generalizations achieved in this area, includes the important progress via the construction of nonlinear entanglement witnesses (NEWs) out of the standard EWs, which are linear over quantum states. They have the property of detecting more entangled states than their linear counterparts.   
Both for standard EWs and NEWs, entanglement detection is guaranteed only when the measurements performed in experiments are ideal. The concept of measurement-device-independent (MDI) EWs - which was then generalized for NEWs - was developed to overcome this limitation. The idea is to use a set of trusted quantum states as inputs, ``quantum inputs", by local observers to detect the present entanglement in a shared state, whereby the witness becomes immune to misalignment in the measurement apparatuses - in the sense that separable states are not erroneously declared as entangled.


In this article, we investigated the performance of MDI-NEWs under two categories of noise - unrelated to misalignments - that may creep in an experiment. In the first category, we studied the effect of detection loophole in the outcomes of the measurements on MDI-NEWs. We considered three cases within this category, \emph{viz.}, detection loophole due to lost events, the same due to additional events, and the same where both the lost and additional events are possible. For each case, we provided the bounds which will guarantee  entanglement in spite of presence of the wrong events. These bounds depend on the efficiencies of the measurement devices, and are defined in terms of the lost, additional, and total events of the outcomes while performing the relevant measurements. In the second category, we considered the instance of noise present in the quantum inputs required for performing an experiment corresponding to a MDI witness. We identified a category of noise, for which MDI-NEWs still guarantees entanglement measurement-device-independently. We also compared the performance of MDI-EWs and MDI-NEWs for noise in quantum inputs. 
It is shown that MDI-NEWs 
are  less or equally robust in comparison to the corresponding MDI-EWs in the presence of noise in quantum inputs, even though the nonlinear ones detect a larger volume of entangled states than their linear kin.
\section*{Acknowledgments}
The research of CS was partly supported  by the INFOSYS scholarship.
We acknowledge partial support from the Department of Science and Technology,
Government of India through the QuEST grant (grant number
DST/ICPS/QUST/Theme-3/2019/120).

\bibliography{apssamp}
\end{document}